\documentclass[letterpaper,10pt,twocolumn]{article}

\usepackage[margin=0.75in]{geometry}

\usepackage{times}
\usepackage{helvet}
\usepackage{courier}

\usepackage[hyphens]{url}
\usepackage{graphicx}
\usepackage[numbers,sort&compress]{natbib}
\usepackage{caption}

\usepackage{amsmath}
\usepackage{amsfonts}
\usepackage{booktabs}
\usepackage[T1]{fontenc}

\usepackage{algorithm}
\usepackage{algorithmic}

\usepackage{newfloat}
\usepackage{listings}

\DeclareCaptionStyle{ruled}{
    labelfont=normalfont,
    labelsep=colon,
    strut=off
}
\floatstyle{ruled}
\newfloat{listing}{tb}{lst}{}
\floatname{listing}{Listing}

\usepackage{xcolor}

\title{
Competing for a Finite Pool of Attention in Social Media?\\
How a New Geopolitical Conflict Reshapes Engagement in Bluesky
}

\author{
Kamand Kalashi,
Arash Badie-Modiri,
Ali Salloum,\\
Juhi Kulshrestha,
Talayeh Aledavood,
Mikko Kivelä\\[0.5em]
\small Department of Computer Science, Aalto University\\
\small Espoo, 00076, Finland\\
\footnotesize \{kamand.kalashi, arash.badie-modiri, ali.salloum, juhi.kulshrestha,\\
\footnotesize talayeh.aledavood, mikko.kivela\}@aalto.fi
}

\date{}

\begin{document}

\maketitle

\begin{abstract}
Major geopolitical crises can rapidly reshape online public attention.
Yet population-level increases in discussion volume about a new crisis
reveal little about how users accommodate this new demand for attention.
We study the onset of the Iran--US--Israel conflict, triggered on
28 February 2026, using longitudinal repost activity from Bluesky
across four consecutive approximately three-month windows spanning the
period before and after its onset; the data comprise 91.0 million unique
posts and 645.5 million repost observations. We find that the new conflict
reorganized participation through both reallocation among existing conflict
participants and substantial activation of previously low-conflict-active
users, while some previously active users reduced their conflict-related
participation. Attention redistribution differed substantially across
pre-existing interests: Iran--US--Israel and Israel--Palestine attention
showed strong positive co-movement with little systematic relative
replacement, whereas Other Political and Non-Political content more
consistently lost attention share, and Russia--Ukraine exhibited weaker,
heterogeneous replacement. Finally, disruption of users' broader attention
allocation was substantially more prevalent among users with established
attention to geopolitical conflicts than in the overall or Non-Political
populations. Together, these findings show that a newly emerging conflict
reorganizes online attention through turnover in who participates, selective
co-attendance or replacement across topics, and disruption of broader
attention patterns concentrated among users already engaged with
geopolitical conflicts.
\end{abstract}


\section{Introduction}
\label{sec:introduction}

Social media platforms have become central arenas for agenda setting,
information sharing, public attention, and political discussion
\cite{neuman2014dynamics,feezell2018agenda,conway2015twitter}. As new
topics emerge, users encounter competing streams of information and must
allocate their limited attention across multiple ongoing events
\cite{falkinger2008limited,feng2015competing}. How users respond to these
competing demands determines which political discussions remain salient
and which gradually recede from the public sphere
\cite{neuman2014dynamics,stewart2020collective}. The collective
allocation of user attention thus constitutes a self-organized form of
gatekeeping that rivals %supplements
the traditional role of journalists and editors in determining which
events reach and remain on the public agenda
\cite{meraz2013networked,shoemaker2020gatekeeping}. These stakes are
particularly high for ongoing crises, such as disasters and conflicts,
where public attention can shape political pressures, funding for
humanitarian aid, and government priorities
\cite{eisensee2007news,olsen2003humanitarian}.

There are seemingly conflicting perspectives on what
happens to existing topics when a new major crisis emerges. Under a strict zero-sum interpretation of finite attention, increased attention to one topic should reduce the attention available for others \cite{zhu1992issue,sisco2023finite,smirnov2022covid}.
% \cite{zhu1992issue}: if attention derives from a finite pool, then an
% increase in one topic requires a decrease in another
% \cite{sisco2023finite,smirnov2022covid}. 
Yet connected issues may also gain attention together
\cite{jang2017redirecting,sisco2023finite,repke2024global}, and observable social media activity can increase as users become more active. Aggregate attention can therefore change through several user-level pathways: users already engaged
in discussion can reallocate their attention, previously inactive users
can become newly active, or previously active users can disengage
\cite{he2017collective,sasahara2013quantifying}. 
Distinguishing these processes is important because the same aggregate increase can reflect either changes in who participates or changes in how existing participants allocate their attention.
% These questions are
% theoretically linked: if individual users' attention cannot grow, then
% aggregate attention can only increase through incoming users. 
Existing
approaches to analyzing aggregate attention patterns cannot distinguish
between these pathways directly. Section~\ref{sec:related-work} reviews this
literature.

We study these questions around the onset of the Iran--US--Israel
conflict on 28 February 2026 \cite{reuters2026iranattack} using
longitudinal repost activity from Bluesky. Bluesky provides a
particularly useful setting for studying attention allocation because,
unlike platforms organized around a single centrally controlled
recommendation algorithm, it gives users substantial control over how
their feeds are curated: users can access a reverse-chronological feed of
accounts they follow and choose among alternative custom feeds rather
than being restricted to a single platform-determined ranking system
\cite{quelle2024bluesky}. Because this design reduces the influence of
any single recommendation mechanism, it makes user-driven attention
allocation across competing topics directly observable. 

We focus on three
geopolitical conflicts that provide a structured comparison rather than
attempting to cover every contemporaneous conflict. Israel--Palestine is
the closest substantive comparison because Israel is a central actor in
both conflicts and the two are embedded in the same broader regional
geopolitical context \cite{kamrava2026iran}; moreover, Israel--Palestine
had already been the subject of sustained attention before the new
conflict emerged \cite{ng2025prominent}. Russia--Ukraine, by contrast, is
another major ongoing conflict but a more distinct substantive context,
providing a less closely related benchmark \cite{mets2026crisis}.
Comparing both allows us to examine how a newly emerging conflict
interacts with a closely related and a more distant pre-existing
conflict, and we extend the comparison beyond conflict-focused discussion
by including \textit{Other Political} and \textit{Non-Political} content.

We operationalize attention through reposting behavior. Reposting
reflects an explicit decision to recirculate existing content and extend
its visibility to a new audience
\cite{metaxas2015retweets,park2019visibility,wang2021attention}, in
contrast to, for example, liking, a lower-commitment interaction that can
express a wide range of reactions \cite{wang2022liking}. Original
posting, in turn, requires content creation and occurs much less
frequently than reposting in our data. Reposts therefore provide a useful
behavioral measure of attention: they capture active engagement and
amplification \cite{metaxas2015retweets,wang2021attention} while
occurring at sufficient volume to characterize how individual users
allocate attention across topics over time.

We go beyond aggregate attention patterns by following the same users
across four approximately three-month windows, two before the conflict (T0, T1), one
covering its onset (T2), and one after (T3). Spanning 91.0 million
unique posts and 645.5 million reposts, this design allows us to ask
how the emergence of the Iran--US--Israel conflict reorganized online
attention at three levels. First, who participates: attention to the new
conflict grew both from existing participants in prior conflicts and
from previously low-conflict-active users. Second, where attention
moves: redistribution was not zero-sum but selective. Israel--Palestine
was co-attended with the new conflict, while the within-user shares of most other topics declined, and this replacement was weaker for Russia--Ukraine. Third, how
much attention patterns are disrupted: disruption was concentrated
among users already engaged with prior conflicts and was partially
reversed by T3.

\section{Related Work and Research Questions}
\label{sec:related-work}

\subsubsection{Aggregate Attention Shifts and the Finite Pool of Attention}
Human attention is limited, making attention a scarce resource that must be
distributed across simultaneously competing issues. This constraint
underlies the Finite Pool of Attention and Finite Pool of Worry perspectives
\cite{sisco2023finite,smirnov2022covid},
which propose that heightened attention or concern directed toward one threat
may reduce the resources available for others. Large external events can
therefore alter the salience of issues already competing for public
attention. Evidence from the COVID-19 pandemic provides a particularly clear
example. Climate-related discussion declined following the onset of the
pandemic across Twitter and news media
\cite{loureiro2021covid,smirnov2022covid,rauchfleisch2023covid,
stoddart2023competing}. Combining self-reported attention with Twitter and
news-media measures, \cite{sisco2023finite} found substantially stronger
evidence for a finite pool of attention than for a finite pool of worry.
\cite{repke2024global} similarly documented a substantial decline in
climate-related attention after the onset of COVID-19, followed by recovery
toward pre-pandemic levels. More broadly, collective attention on social
media is often characterized through aggregate quantities such as the
frequency with which an issue or event appears in public discourse
\cite{stewart2020collective}. %These studies establish that major crises can redirect aggregate attention, but aggregate change alone does not identify the user-level behaviors through which that change occurs.

\subsubsection{User-Level Pathways: Reallocation, Activation, and Contraction}
When a major new crisis emerges, the attention it attracts can be accommodated through several distinct changes in user behavior \cite{he2017collective,zhang2021reddit}. 
Some users
who were already engaged with ongoing conflicts may reallocate their
conflict-related attention toward the new event, changing which conflicts
dominate their activity or combining the new conflict with existing
interests \cite{he2017collective,zhang2021reddit}. At the same time, growth
in discussion of the new crisis need not come only from users who were
already participating in conflict-related discourse. The event may activate
users who previously devoted little or no attention to conflict-related
content, while some previously active users may reduce their
conflict-related participation \cite{zhang2021reddit}. The same aggregate
increase in attention to a new crisis can therefore arise from very different
user-level pathways: reallocation among existing participants, entry of
previously low-conflict-active users, and contraction among others
\cite{zhang2021reddit}. Tracing these movements is necessary to understand
not only how much attention a new conflict receives, but how the population
participating in conflict discussion is reorganized around it
\cite{he2017collective,zhang2021reddit}.

Large events can alter not only what is discussed but also who participates
and how users behave. During COVID-19, information-seeking behavior on
Wikipedia underwent substantial changes in both volume and topical
composition \cite{hortaribeiro2021covid}, while online communication
patterns on Snapchat changed across public and private modes of interaction
\cite{yang2021communication}. A longitudinal analysis of Twitter likewise
examined changes in posting activity alongside changes in the prominence of
different topics around the onset of the pandemic \cite{valdez2020social}.
At the community level, \cite{zhang2021reddit} documented substantial
movement of users between closely related COVID-19 communities as the
pandemic unfolded. 
%These findings show why population-level increases or declines cannot by themselves distinguish different behavioral pathways. Increased attention to a new crisis may involve users reallocating attention they were already devoting to related discussion, previously low-participation users becoming active, or previously active users reducing their participation. 
%This distinction motivates our first question, which examines reallocation, activation, and contraction directly at the user level rather than inferring them from aggregate changes in issue prevalence.

\subsubsection{Selective and Heterogeneous Redistribution}
Beyond these changes in participation, a new crisis may also reorganize what
users attend to and the broader structure of their attention
\cite{he2017collective,sasahara2013quantifying}. Increased
attention to a new conflict may coincide with an absolute decline in an
existing topic, a reduction only in that topic's relative share of attention,
or positive co-movement in which both topics receive more activity
\cite{jang2017redirecting,sisco2023finite,repke2024global}.
%Importantly, these relationships need not be uniform across topics.
%\cite{sisco2023finite,repke2024global}. 
%Substantively related conflicts may compete more strongly because they draw on overlapping political attention, but the same overlap may instead facilitate co-attention if users interested in one conflict are also likely to engage with the other \cite{sisco2023finite,rauchfleisch2023covid}. 
Attention redistribution also need not take the form of uniform competition
across topics. \cite{sisco2023finite}, for example, found that greater
COVID-19 attention was associated with reduced attention to climate change
and terrorism but increased attention to directly connected economic
problems and unemployment. Similarly, some climate activists explicitly
connected COVID-19 and climate change rather than treating them as
independent issues \cite{rauchfleisch2023covid}, and the pandemic affected
different themes within climate discourse differently
\cite{repke2024global}. Research on crisis communication further shows that
the content and organization of online discussion can change substantially
as events unfold. During the 2009 Israel--Gaza conflict, Twitter users
selectively redistributed information from a concentrated set of sources,
with the prominence of different information channels changing over the
news cycle \cite{kwon2012audience}. Following the 2016 Berlin terrorist
attack, the topical composition of Twitter discussion shifted from
emotion-related and operational content toward more opinion-oriented
discussion over subsequent days \cite{fischer2019collective}. More recent
work on the 2023 Israel--Hamas escalation has similarly documented sharp
event-related peaks in social-media activity and changes in the content
driving engagement across Twitter, Reddit, and TikTok
\cite{ng2025prominent}. Longitudinal work on Ukraine has further shown how
attention can be evaluated relative to prior baseline frequencies, revealing
distinct crisis-related increases around the 2014 and 2022 invasions across
different language communities \cite{mets2026crisis}. Together, this work
demonstrates that crisis-related attention is heterogeneous across issues,
content, audiences, and time. What remains less clear is whether a newly
emerging crisis systematically replaces some pre-existing topics more than
others at the level of individual users, particularly whether substantively
related topics compete for overlapping attention or are instead co-attended
by users with shared interests.

\subsubsection{The Structure of Attention Allocation}
Topic-specific changes,
however, capture only part of the potential disruption. A major event can
alter the composition of a user's attention, changing across several categories
at once \cite{he2017collective,sasahara2013quantifying,
purohit2014understanding}. 
%We distinguish between where attention is redistributed and how extensively the user's overall attention allocation pattern is reorganized. By comparing change during the conflict transition with change across an earlier pre-conflict transition, we can assess whether the event period represents greater disruption than the user's preceding level of temporal variation, which category-level shifts contribute to that disruption, and whether users subsequently return toward their pre-conflict attention patterns or remain displaced from them.
A complementary perspective concerns the structure of attention allocation
itself. Rather than representing attention only through the amount devoted
to a single issue, behavior can be characterized by how an individual
divides attention among multiple alternatives. On social media, this has been conceptualized as an ``information diet'', i.e., the topical distribution of information a user produces or consumes \cite{kulshrestha2015characterizing}.
Prior work in other domains has found that individuals allocate attention across targets in stable patterns even as the targets themselves change \cite{backstrom2011center,saramaki2014persistence,baltakys2023investor}.

\subsection*{Research Questions}

Aggregate changes in attention do not by themselves reveal the user-level
processes that produce them. We therefore first examine how user-level
participation dynamics, specifically changes in who participates and which
conflicts they attend to, account for the aggregate rise in attention
surrounding the new conflict (RQ1). We then examine how attention is
redistributed and reorganized within individual users, independent of any
particular aggregate outcome (RQ2).

\begin{enumerate}

\item \textbf{RQ1 (Reallocation, Activation, and Contraction):}
\textit{What user-level participation dynamics underlie the aggregate increase
in attention created by the emergence of a major new conflict?}

We examine whether increased attention to the Iran--US--Israel conflict is
accommodated by reallocating attention that users were already devoting to
other conflicts. We also ask whether the event draws previously less-engaged
users into conflict-related discussion, while some previously active users
reduce their participation. Finally, we examine whether these changes persist
after the initial conflict period or whether users return toward their
pre-conflict patterns of conflict attention.

\item \textbf{RQ2 (Attention Redistribution and Disruption):}
\textit{How does the emergence of a major new conflict reorganize both the
distribution and the structure of attention within individual users?}

\textbf{RQ2a (Redistribution and Selective Competition):}
How is attention redistributed after the emergence of the
Iran--US--Israel conflict, and how does this redistribution differ across
users' pre-existing interests? In particular, for which existing interests
is increased attention to the new conflict accommodated through an
expansion of users' overall attention pool rather than by drawing from
attention previously devoted to those interests, and for which does the
new conflict more strongly replace existing attention?

\textbf{RQ2b (Disruption of Attention Patterns):}
To what extent does the emergence of the Iran--US--Israel conflict disrupt
users' characteristic patterns of attention allocation across content
categories, relative to their pre-conflict level of change? Which shifts in
attention contribute most strongly to this disruption, and do users
subsequently return toward their pre-conflict attention patterns, or does the
disruption persist?

\end{enumerate}

\section{Data}
\label{sec:data}

\begin{figure*}[!t]
\centering
\includegraphics[width=0.8\textwidth]{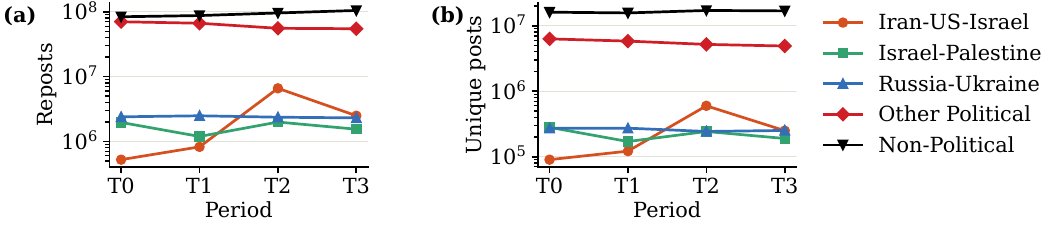}
\caption{Repost observations and unique posts across content categories and
observation windows. Panel (a) shows the number of repost observations and
Panel (b) shows the number of unique posts for each content category across
T0--T3. The y-axes use logarithmic scales.}
\label{fig:topic-distribution}
\end{figure*}

We collected public activity from the Bluesky social network using the
Sinitaivas Live collection pipeline \cite{sinitaivas2025live}. We analyze
four approximately three-month windows surrounding the emergence of the
Iran--US--Israel conflict on 28 February 2026. T0
(1 September--29 November 2025) provides an earlier reference period, T1
(30 November 2025--27 February 2026) captures the period immediately preceding
the conflict, T2 (28 February--28 May 2026) captures the conflict period, and
T3 (29 May--26 August 2026) captures the subsequent three-month period.
We exclusively analyze repost events of English-language posts that received
at least one repost. Figure~\ref{fig:topic-distribution} shows the
distribution of repost observations and unique posts across categories and
observation windows. %with exact counts and category shares reported in
%Appendix~\ref{app:collection}, 
%Table~\ref{tab:topic-distribution} (details on data collection, filtering, and coverage in
%Appendix~\ref{app:collection}).

We classified posts using Gemma-4-26B-A4B-it into five mutually exclusive
categories: Iran--US--Israel Conflict, Israel--Palestine Conflict,
Russia--Ukraine Conflict, Other Political, and Non-Political. Classification
used a structured-output prompt with deterministic generation settings
(temperature = 0) (complete classification prompt and category
definitions in Appendix~\ref{app:classification-prompt}).

We assessed the reliability of the LLM-generated topic labels against manual
coding. We drew a stratified uniformly random validation sample from the set of all
unique posts in T0--T3, sampling separately from each LLM-assigned topic
category, and manually coded the 239 randomly selected posts from this subset
while blinded to the LLM-assigned labels. Agreement was high across categories:
the macro-average class-specific precision was 95.4\% (95\% bootstrap CI:
92.4--97.9\%). Because the validation sample deliberately over-sampled the
less common topic categories, we additionally reweighted the validation results
using the distribution of LLM labels in the full corpus. The resulting estimated
corpus-wide classification accuracy was 97.1\% (95\% stratified-bootstrap CI:
92.7--99.8\%). Class-specific precision ranged from 86.8\% for the
Iran--US--Israel category to 100\% for the Israel--Palestine category (details of the validation procedure and category-level results
in Appendix~\ref{app:llm-validation}).

\section{Behavioral Responses to the Emerging Conflict (RQ1)}
\label{sec:rq1}

RQ1 examines how users accommodated the increased demand for attention
associated with the emergence of the Iran--US--Israel conflict. At the
aggregate level, Figure~\ref{fig:topic-distribution}
%Table~\ref{tab:topic-distribution} 
shows that
Iran--US--Israel attention rose sharply during T2 before declining in T3,
while Israel--Palestine attention declined in T1, rebounded in T2, and
declined again in T3. Russia--Ukraine remained comparatively stable, whereas
Other Political content declined across the observation windows, as
Non-Political content became increasingly prominent. We first
characterize how users distributed their conflict-related attention and trace
changes in these allocations from T1 to T2 to T3. These transitions allow us to
distinguish reallocation among existing conflict topics from the activation of
previously low-conflict-active users and contraction among previously active
participants. We then examine whether these individual-level changes were
accompanied by a broader reorganization of conflict-focused audiences in the
repost network.

\subsection{Conflict-Attention Profiles and User Transitions}
\label{sec:rq1_profiles}

\subsubsection{Constructing Conflict-Attention Profiles}
\label{sec:conflict_profiles}

To characterize how users allocated attention specifically within
conflict-related discussion, we constructed  \textit{conflict-attention profiles} based
on repost activity concerning the three focal conflicts: Iran--US--Israel
(\textit{IUI}), Israel--Palestine (\textit{IP}), and Russia--Ukraine
(\textit{RU}). For each user $u$ and time window $t$, we counted the number
of distinct posts reposted from each conflict and normalized these counts by
the user's total number of conflict-related reposted posts in that window.
The resulting conflict attention profile is
\[
\mathbf{c}_{u,t}
=
\left(
\frac{n_{u,t}^{IUI}}{N_{u,t}^{C}},
\frac{n_{u,t}^{IP}}{N_{u,t}^{C}},
\frac{n_{u,t}^{RU}}{N_{u,t}^{C}}
\right),
\]
where $n_{u,t}^{k}$ is the number of distinct posts from conflict $k$
reposted by user $u$ during window $t$, $C$ denotes the set of the three
focal conflict categories, and
%\[
$N_{u,t}^{C}
=
n_{u,t}^{IUI}
+
n_{u,t}^{IP}
+
n_{u,t}^{RU}$.
%\]
Thus, among users with some conflict-related activity, the three components sum
to one and describe the relative allocation of a user's conflict attention,
independently of their non-conflict repost activity.

For analyses based on conflict-attention profiles, we retained users who
reposted at least five distinct posts belonging to any of the three focal
conflicts across the observation windows. Within each window, users with
fewer than three conflict-related reposts were treated as having insufficient
activity for profile classification. The remaining users were assigned to a
dominant-conflict profile when at least 60\% of their conflict attention was
allocated to one conflict, a two-conflict profile when each of two conflicts
accounted for at least 25\%, or an all-conflict profile when each of the three
conflicts accounted for at least 20\%. Remaining combinations were
classified as \textit{Other Mixed}.

\subsubsection{Profile Transitions, Activation, and Contraction}
\label{sec:profile_transitions}

We traced changes in users' conflict-attention profiles from the pre-conflict
window (T1), through the first three months following the conflict onset (T2), to the subsequent three-month window (T3), and visualized these transitions using
an alluvial diagram (Fig.~\ref{fig:conflict_signatures}(a)).
Transitions between classified profiles indicate reallocation of
conflict-related attention. Movement from insufficient activity into a
classified profile indicates activation, whereas movement in the opposite
direction indicates contraction. Extending the transitions through T3
further reveals whether changes observed following conflict onset persisted
or whether users returned toward their previous conflict-attention profiles.

The emergence of the Iran--US--Israel conflict produced a sharp reorganization
of conflict attention (Fig.~\ref{fig:conflict_signatures}(a)). The
Iran--US--Israel-dominant profile expanded more than tenfold from T1 to T2,
drawing particularly strongly from users previously focused on
Russia--Ukraine: nearly half of Russia--Ukraine-dominant users shifted
directly to an Iran--US--Israel-dominant profile. Israel--Palestine-focused
users were less likely to make this direct switch and more often retained
Israel--Palestine attention, either as their dominant profile or alongside
Iran--US--Israel. At the same time, the All Three Conflicts profile contracted
substantially, indicating that the initial response to the emerging conflict
concentrated attention around Iran--US--Israel rather than broadening attention
uniformly across conflicts. The shift also involved substantial activation of
users with previously insufficient conflict activity, alongside some
contraction among previously active users.

By T3, this initial concentration had partially reversed. The
Iran--US--Israel-dominant population fell substantially from its T2 peak,
while both the Russia--Ukraine- and Israel--Palestine-dominant populations
rebounded. The return was nevertheless incomplete: Iran--US--Israel remained
considerably more prominent than before conflict onset, and many users
continued to allocate attention across Iran--US--Israel and the pre-existing
conflicts. Together, the transitions suggest a pronounced initial shift toward
the emerging conflict followed by partial return toward earlier
conflict-attention patterns, with Russia--Ukraine-focused users showing a
stronger initial replacement of their prior focus than Israel--Palestine-focused
users.

\subsection{Reorganization of Conflict-Focused Audiences}
\label{sec:rq1_network}

To examine whether changes in conflict attention were accompanied by changes
in audience organization, we constructed a directed, weighted user--user
repost network for each window, where nodes represent users and edges represent
reposting of another user's conflict-related content. Edge weights count the
number of distinct conflict-related posts underlying each relationship, and
we retained edges supported by at least five such posts. Nodes were linked to
their conflict-attention profile for the corresponding window
(Section~\ref{sec:conflict_profiles}), allowing changes in audience structure
to be compared across T1, T2, and T3. We additionally calculated, for each
source profile, the proportion of its outgoing repost links directed toward
each target profile; the resulting row-normalized mixing matrices are reported
in Appendix~\ref{app:profile-mixing}.

\begin{figure*}[!t]
\centering
\includegraphics[width=0.85\textwidth]{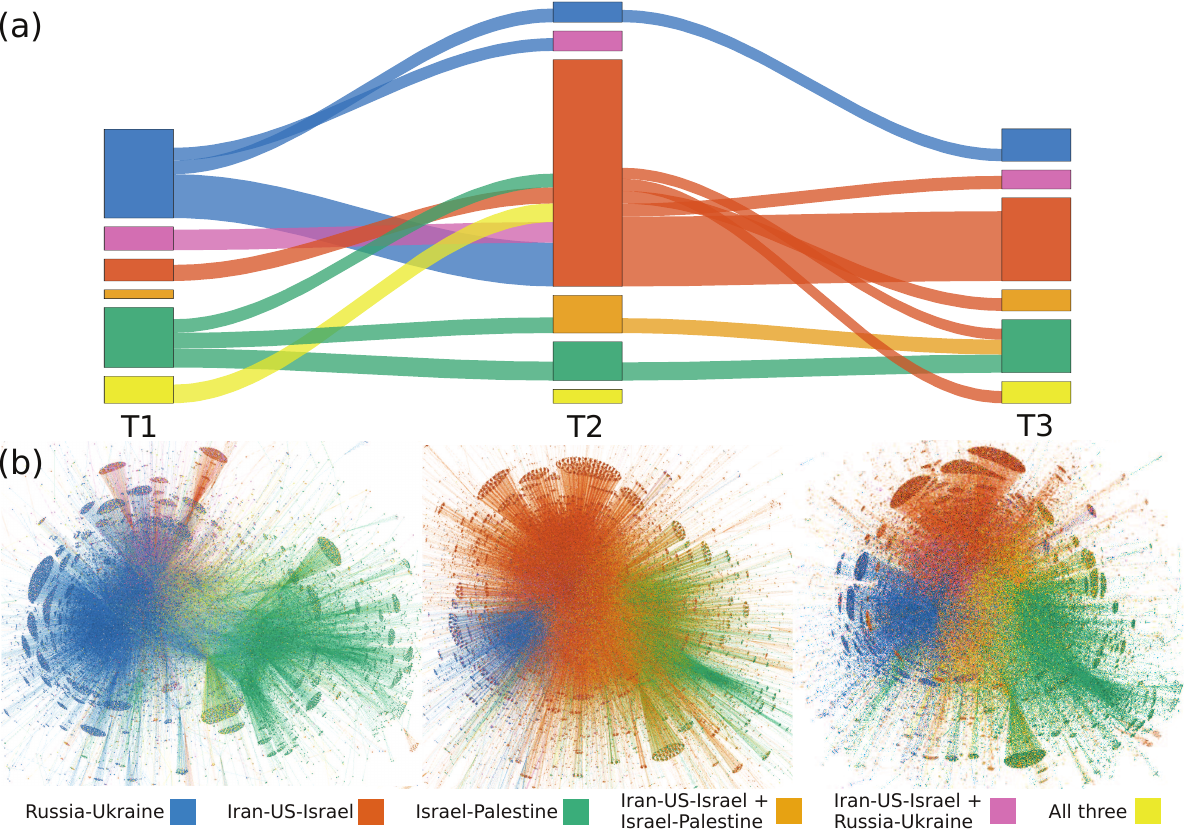}
\caption{Changes in conflict-attention profiles and conflict-focused repost
networks across T1--T3. (a) Alluvial diagram of transitions between conflict-attention profiles; node sizes represent profile populations and flow widths represent the number of transitioning users. For visual clarity, only profile classes and flows containing at least 5,000 users are shown.
(b) Directed repost networks of
conflict-related interactions in T1, T2, and T3. Nodes represent users,
edges represent repost relationships, and colors indicate users'
conflict-attention profiles using the same scheme across panels.}
\label{fig:conflict_signatures}
\end{figure*}

The networks reveal a pronounced reorganization of conflict-focused audiences
(Fig.~\ref{fig:conflict_signatures}(b)). In T1, Russia--Ukraine and
Israel--Palestine focused users occupied relatively separated regions of the
network. With the emergence of the Iran--US--Israel conflict in T2, this
structure became substantially more integrated around a large
Iran--US--Israel-focused core connecting previously differentiated audiences.
The mixing matrices support this visual pattern (Table~\ref{tab:profile-mixing-matrices} in Appendix~\ref{app:profile-mixing}):
the share of outgoing links remaining within the Russia--Ukraine profile fell
in T2, while a greater share was directed toward Iran--US--Israel users.
Among Israel--Palestine users, within-profile linking similarly declined,
while links toward the combined Iran--US--Israel+Israel--Palestine profile
became substantially more prominent. By T3, within-profile linking rebounded
for both Russia--Ukraine and Israel--Palestine, consistent with the partial
return toward earlier attention profiles observed in the alluvial diagram,
but the network did not return to its T1 structure. Instead, links involving
Iran--US--Israel remained more prominent than before the conflict, suggesting
that the emerging conflict not only redirected individual attention but also
increased connections across previously more differentiated conflict audiences.

\section{Attention Redistribution and Disruption (RQ2)}
\label{sec:rq2}

RQ2 examines how the emergence of the Iran--US--Israel conflict reorganized
both the distribution and broader structure of users' attention. We first
examine how increased attention to the new conflict was accommodated across
users' pre-existing interests, distinguishing expansion of overall attention
from replacement of attention previously devoted to those interests (RQ2a).
We then examine whether the conflict disrupted users' broader patterns of
attention allocation relative to their pre-conflict level of change, which
shifts contributed most strongly to this disruption, and whether users
subsequently returned toward their pre-conflict patterns or the disruption
persisted (RQ2b).

\subsection{Redistribution and Selective Competition (RQ2a)}
\label{sec:rq2a}

\subsubsection{Measuring Changes in Attention}
\label{sec:attention_deltas}

To quantify how each user's attention to a content category changed following
the onset of the Iran--US--Israel conflict, we measured the difference
between the attention devoted to that category in T1 and T2, which we refer
to as an attention change. Let $n_{u,t}^{k}$ denote the number of annotated
reposts by user $u$ belonging to category $k$ in window $t$, where
$k \in \{\mathrm{IUI}, \mathrm{IP}, \mathrm{RU}, \mathrm{OP},
\mathrm{NP}\}$ denotes Iran--US--Israel (IUI), Israel--Palestine (IP),
Russia--Ukraine (RU), Other Political (OP), and Non-Political (NP),
respectively.

We represent attention change in two complementary ways. The
absolute attention change measures the change in the number of reposts devoted
to category $k$:
%\[
$\Delta_{u,k}^{\mathrm{abs}}
=
n_{u,T2}^{k} - n_{u,T1}^{k}.$
%\]
The relative attention change measures how the share of a user's
attention devoted to category $k$ changed between windows. We define the
relative attention allocated to category $k$ as
%\[
$a_{u,t}^{k}
=
\frac{n_{u,t}^{k}}{N_{u,t}},
$
%\]
where $N_{u,t}$ denotes the user's total annotated repost activity in window
$t$, and its change as
\[
\Delta_{u,k}^{\mathrm{rel}}
=
a_{u,T2}^{k} - a_{u,T1}^{k}.
\]
%Positive values indicate that the category occupies a larger share of the
%user's attention in T2 than in T1, whereas negative values indicate a smaller
%share.

Together, the absolute and relative attention changes distinguish changes in
the amount of attention from changes in how users allocate their attention.
For example, a user may have a positive absolute change but little or no
positive relative attention change if their overall repost activity increases at a
similar or faster rate. Conversely, increased absolute attention to the new
conflict without a corresponding decrease in the relative attention devoted
to an existing topic suggests expansion of the user's overall attention pool
rather than replacement of attention previously devoted to that topic.

Our primary analyses compare changes in Iran--US--Israel attention with
changes in Israel--Palestine, Russia--Ukraine, Other Political, and
Non-Political attention. We additionally use all non-IUI content as a
baseline for assessing overall changes in the attention space irrespective
of topic; details of this baseline analysis are provided in
Appendix~\ref{app:non_iui_baseline}.

To ensure that relative attention was defined in both windows and that each
comparison concerned a pre-existing interest, users had to be active in both
T1 and T2 and have at least three reposts in the comparison category across
the two pre-conflict windows:
\[
N_{u,T1} > 0,\qquad
N_{u,T2} > 0,\qquad
n_{u,T0}^{k} + n_{u,T1}^{k} \geq 3.
\]
The category-specific requirement applies to Israel--Palestine,
Russia--Ukraine, Other Political, and Non-Political comparisons.

Within each comparison population, we further stratified users by their T1
activity and political engagement to examine whether attention redistribution
varied with prior behavior. Total T1 repost activity was divided into five
levels: 1--4, 5--14, 15--34, 35--69, and at least 70 reposts. We defined
political engagement as the share of T1 attention devoted to political
content:
$p_{u,T1}^{\mathrm{pol}}
=
(n_{u,T1}^{\mathrm{IUI}}
+
n_{u,T1}^{\mathrm{IP}}
+
n_{u,T1}^{\mathrm{RU}}
+
n_{u,T1}^{\mathrm{OP}}
)/
N_{u,T1}$,
and divided it into three levels: $[0,0.3)$, $[0.3,0.6)$, and $[0.6,1]$.
Crossing the five activity levels with the three political-engagement levels
produced 15 behavioral groups, allowing us to assess whether redistribution
differed by users' prior activity and political engagement.

\subsubsection{Relationships Between Changes in Attention}
\label{sec:bivariate_deltas}

We examined the user-level relationship between changes in
Iran--US--Israel attention and changes in each pre-existing interest. For
each comparison category $k$, we jointly analyzed users' absolute attention
changes,
%\[
$
\left(
\Delta_{u,\mathrm{IUI}}^{\mathrm{abs}},
\Delta_{u,k}^{\mathrm{abs}}
\right)$,
%\]
and, separately, their relative attention changes,
%\[
$\left(
\Delta_{u,\mathrm{IUI}}^{\mathrm{rel}},
\Delta_{u,k}^{\mathrm{rel}}
\right)$.
%\]

These bivariate distributions show whether user-level changes in attention
to the emerging conflict coincide with increases, decreases, or little
change in attention to each existing topic. 
Within each of the 15 behavioral groups, we computed Pearson's $r$ between the two attention changes, enabling comparison across prior activity and political engagement (results
for the non-IUI baseline in
Appendix~\ref{app:non_iui_baseline}).

\subsubsection{Attention Expansion and Replacement}
\label{sec:bivariate_redistribution_results}

\begin{figure*}[!t]
\centering
\includegraphics[width=0.8\textwidth]
{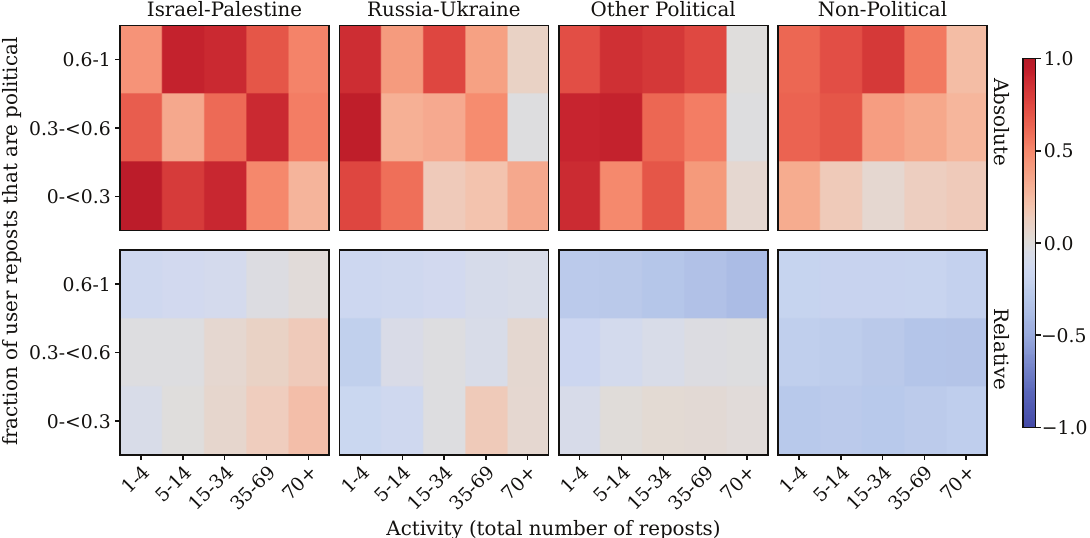}
\caption{User-level relationships between changes in Iran--US--Israel
attention and Israel--Palestine, Russia--Ukraine, Other Political, and
Non-Political attention across T1 behavioral groups. Heatmaps show Pearson
correlations between the two attention changes, with absolute attention
changes in the top row and relative attention changes in the bottom row.
Within each heatmap, columns represent users' total annotated repost activity
in T1 (1--4, 5--14, 15--34, 35--69, and 70+ reposts), and rows represent
the share of their T1 attention devoted to political content
($[0,0.3)$, $[0.3,0.6)$, and $[0.6,1]$).}
\label{fig:attention-delta-correlations}
\end{figure*}

Figure~\ref{fig:attention-delta-correlations} summarizes the relationships
between changes in Iran--US--Israel attention and each pre-existing interest
across the 15 behavioral groups. Absolute attention generally moved in the
same direction across topics: users who increased their attention to
Iran--US--Israel also tended to increase their attention to existing topics.
For Israel--Palestine, this positive co-movement was strong across both
lower- and higher-activity users. For Other Political content, it was pronounced among lower-activity users with moderate to high
prior political engagement. Russia--Ukraine showed weaker co-movement among
highly active and politically engaged users, while Non-Political content
generally exhibited the weakest positive association. Overall, the emerging
conflict was often accommodated through increased activity rather
than an absolute reduction in attention to existing topics.

Relative attention reveals a different picture. As Iran--US--Israel occupied
a larger share of users' attention, the shares devoted to Other Political and
Non-Political content generally declined, indicating relative replacement
despite increases in their absolute activity. Russia--Ukraine showed weaker
but similar evidence of replacement, particularly among highly politically
engaged and low activity users.

Israel--Palestine was the clear exception: attention to it increased strongly
alongside Iran--US--Israel in absolute terms while its relative share showed
little systematic decline. Thus, increased attention to the emerging
conflict was accommodated differently across pre-existing interests.
Israel--Palestine was largely co-attended through an expansion of overall
attention, whereas Russia--Ukraine and especially broader political and
non-political interests showed greater evidence of relative replacement 
(population sizes and correlation coefficients and results for the non-IUI baseline in
Appendix~\ref{app:non_iui_baseline}).

\subsubsection{Conditional Relationships Between Changes in Attention}
\label{sec:multivariate_method}

We next used multiple linear regression to examine how changes in each
pre-existing interest were independently associated with changes in
Iran--US--Israel attention when considered simultaneously. The analysis
extends the absolute attention change, $\Delta_{u,k}^{\mathrm{abs}}$, defined
above to both the conflict-period transition (T1--T2) and the preceding
transition (T0--T1), and includes only users active in all three windows.

The dependent variable was the T1--T2 absolute Iran--US--Israel attention
change. Predictors included the T1--T2 absolute changes for Israel--Palestine,
Russia--Ukraine, Other Political, and Non-Political attention, together with
their corresponding T0--T1 attention changes. Including the preceding-period changes
allows the relationships observed during the conflict transition to be
estimated while accounting for users' recent changes in the same categories.

We estimated the model using ordinary least squares on the original
repost-count scale. Each coefficient therefore represents the change in the
T1--T2 absolute Iran--US--Israel attention change associated with a one-repost
increase in the corresponding predictor change, holding the other predictors
constant. We used HC3 heteroskedasticity-robust standard errors for inference
\cite{mackinnon1985heteroskedasticity}.

\subsubsection{Topic-Specific Conditional Associations}
\label{sec:multivariate_results}

The multivariate model included 986,288 users active in all three observation
windows and explained 31.1\% of the variation in the T1--T2 absolute
Iran--US--Israel attention change ($R^2=.311$, adjusted $R^2=.311$;
Table~\ref{tab:absolute-multiple-regression}). Thus, changes in the included
content categories and their preceding-period changes jointly accounted for
a substantial portion of the variation in users' increased or decreased
Iran--US--Israel activity, while approximately 70\% of the variation remained
unexplained by changes in the other topics, providing evidence for substantial
independence in how attention to different topics changed.

Among the T1--T2 relationships, three were statistically significant.
Israel--Palestine showed by far the strongest positive relationship with
Iran--US--Israel ($\beta=2.166$, $p<.001$). Holding the other predictors
constant, a one-repost larger increase in Israel--Palestine attention was
associated with a 2.166-repost larger increase in Iran--US--Israel attention.
Russia--Ukraine change was also positively associated with Iran--US--Israel
change ($\beta=.196$, $p=.032$), whereas Other Political change showed a
small negative association ($\beta=-.046$, $p<.001$). Non-Political change
was not statistically significant ($p=.068$). These results reinforce the
bivariate finding that Israel--Palestine was primarily co-attended with the
emerging conflict rather than replaced by it, while Other Political attention
showed limited conditional competition in absolute activity.

The preceding T0--T1 changes provided additional context rather than the
main substantive result. Prior Israel--Palestine change was negatively
associated with subsequent Iran--US--Israel change ($\beta=-.382$,
$p<.001$), whereas prior Russia--Ukraine and Other Political changes showed
positive associations; prior Non-Political change was not statistically
significant.

\begin{table}[t]
\centering
\setlength{\tabcolsep}{3pt}

\begin{tabular}{lrrr}
\hline
Predictor & $\beta$ & 95\% CI & $p$ \\
\hline
\multicolumn{4}{l}{\textit{T1--T2 changes}} \\
Other Political
    & $-0.046$ & $[-0.062,\,-0.030]$ & $<.001$ \\
Israel--Palestine
    & $2.166$ & $[1.679,\,2.653]$ & $<.001$ \\
Russia--Ukraine
    & $0.196$ & $[0.017,\,0.375]$ & $.032$ \\
Non-Political
    & $0.013$ & $[-0.001,\,0.027]$ & $.068$ \\[2pt]

\multicolumn{4}{l}{\textit{T0--T1 changes}} \\
Other Political
    & $0.015$ & $[0.0005,\,0.029]$ & $.043$ \\
Israel--Palestine
    & $-0.382$ & $[-0.607,\,-0.157]$ & $<.001$ \\
Russia--Ukraine
    & $0.146$ & $[0.062,\,0.230]$ & $<.001$ \\
Non-Political
    & $0.002$ & $[-0.008,\,0.011]$ & $.749$ \\
\hline
\multicolumn{4}{l}{
$n=986{,}288$;\quad
$R^2=.311$;\quad
adjusted $R^2=.311$.}
\end{tabular}

\caption{Multiple OLS regression of the absolute Iran--US--Israel
attention change from T1 to T2. Coefficients are reported in repost units.
Confidence intervals and $p$-values use HC3 heteroskedasticity-robust
inference.}
\label{tab:absolute-multiple-regression}
\end{table}

\subsection{Disruption of Attention Patterns (RQ2b)}
\label{sec:rq2b}

RQ2a examined whether changes in Iran--US--Israel attention were accompanied
by changes in particular pre-existing interests. RQ2b instead considers each
user's full five-category attention distribution as a single structure. We
ask whether this structure changed unusually strongly around conflict onset
relative to the user's own pre-conflict change, which category-level shifts
characterized this reorganization, and whether users subsequently returned
toward their immediately pre-conflict attention pattern or remained
displaced from it.

\subsubsection{Measuring Disruption, Category Shifts, and Return}
\label{sec:attention_disruption_method}

Using the relative attention values defined in
Section~\ref{sec:attention_deltas}, we represent user $u$'s attention in
window $t$ as
\[
\mathbf{a}_{u,t}
=
\left(
a_{u,t}^{RU},
a_{u,t}^{IP},
a_{u,t}^{OP},
a_{u,t}^{NP},
a_{u,t}^{IUI}
\right).
\]
We restrict the analysis to users active in all four windows, allowing the
same users to be followed from the pre-conflict period through T3. The
Israel--Palestine, Russia--Ukraine, Other Political, and Non-Political
populations use the same prior-attention criterion defined in
Section~\ref{sec:attention_deltas}; these populations may overlap.

We measure differences between attention distributions using
Jensen--Shannon divergence (JSD). 
JSD is zero for identical attention
distributions and increases as their compositions diverge. For each user, we
define
\[
\begin{aligned}
d_{u,ab}
&=
\operatorname{JSD}
\left(
\mathbf{a}_{u,T_a},
\mathbf{a}_{u,T_b}
\right),
\qquad ab\in\{01,12,13\},\\
\Delta d_u^{\mathrm{dis}}
&=
d_{u,12}-d_{u,01},
\qquad
\Delta d_u^{\mathrm{ret}}
=
d_{u,13}-d_{u,12}.
\end{aligned}
\]
Here, $\Delta d_u^{\mathrm{dis}}>0$ indicates greater attention-pattern
change across conflict onset than during the preceding pre-conflict
transition. Conversely, $\Delta d_u^{\mathrm{ret}}<0$ indicates that the
user's T3 attention distribution is closer to their immediately pre-conflict
T1 distribution than their T2 distribution was, whereas
$\Delta d_u^{\mathrm{ret}}>0$ indicates persistence or further divergence.
For each population, we report the mean JSDs and contrasts together with the
fractions of users satisfying $\Delta d_u^{\mathrm{dis}}>0$ and
$\Delta d_u^{\mathrm{ret}}<0$, expressed as percentages.

To identify which category-level movements accompanied this structural
change, we extend the relative-attention change defined in
Section~\ref{sec:attention_deltas} to the T0--T1, T1--T2, and T2--T3
transitions. We denote the population mean of this user-level change for
category $k$ over transition $T_a$--$T_b$ by
$\overline{\Delta}^{\mathrm{rel}}_{k,ab}$ and report it in percentage
points.

\subsubsection{Attention-Pattern Disruption and Return}
\label{sec:attention_disruption_results}

\begin{table*}[!t]
\centering
\small
\setlength{\tabcolsep}{1mm}

\begin{tabular}{@{}lrrrrrrrr@{}}
\hline
\textbf{Population}
& $\mathbf{n}$
& $\overline{\mathrm{JSD}}_{01}$
& $\overline{\mathrm{JSD}}_{12}$
& $\overline{\Delta d}^{\mathrm{dis}}$
& $\mathbf{\%\ \Delta d_u^{\mathrm{dis}}>0}$
& $\overline{\mathrm{JSD}}_{13}$
& $\overline{\Delta d}^{\mathrm{ret}}$
& $\mathbf{\%\ \Delta d_u^{\mathrm{ret}}<0}$ \\
\hline

Overall
& 849,032
& .0361
& .0424
& +.0063
& 42.6
& .0427
& +.0003
& 37.0 \\

Israel--Palestine
& 99,022
& .0176
& .0298
& +.0122
& 77.5
& .0264
& -.0033
& 66.3 \\

Russia--Ukraine
& 83,866
& .0165
& .0344
& +.0178
& 86.1
& .0299
& -.0044
& 72.8 \\

Other Political
& 477,583
& .0375
& .0520
& +.0144
& 64.0
& .0529
& +.0009
& 53.3 \\

Non-Political
& 792,156
& .0263
& .0334
& +.0071
& 43.1
& .0336
& +.0001
& 37.2 \\
\hline
\end{tabular}

\caption{Within-user attention-pattern disruption and return toward the
immediately pre-conflict T1 pattern. Overlines denote population means.
For user $u$,
$\Delta d_u^{\mathrm{dis}}=\mathrm{JSD}_{12,u}-\mathrm{JSD}_{01,u}$
measures conflict-period disruption relative to pre-conflict change, whereas
$\Delta d_u^{\mathrm{ret}}=\mathrm{JSD}_{13,u}-\mathrm{JSD}_{12,u}$
measures change in distance from the T1 reference, with
$\Delta d_u^{\mathrm{ret}}<0$ indicating return toward the T1 pattern.
Category-specific populations may overlap.}
\label{tab:attention-pattern-disruption-return}
\end{table*}

\begin{table}[t]
\centering

{\small
\setlength{\tabcolsep}{1mm}

\begin{tabular}{@{}llrrrrr@{}}
\hline
\textbf{Pop.}
& \textbf{Transition}
& \multicolumn{5}{c}{\textbf{Mean relative-attention change (pp)}} \\
\cline{3-7}
&
& $\overline{\Delta}^{\mathrm{rel}}_{RU}$
& $\overline{\Delta}^{\mathrm{rel}}_{IP}$
& $\overline{\Delta}^{\mathrm{rel}}_{OP}$
& $\overline{\Delta}^{\mathrm{rel}}_{NP}$
& $\overline{\Delta}^{\mathrm{rel}}_{IUI}$ \\
\hline

All
& T0$\rightarrow$T1
& +0.01
& -0.54
& +0.12
& +0.30
& +0.11 \\

&
T1$\rightarrow$T2
& -0.08
& +0.30
& \textbf{-3.90}
& +1.50
& \textbf{+2.18} \\

&
T2$\rightarrow$T3
& -0.02
& -0.17
& -0.51
& \textbf{+2.43}
& \textbf{-1.73} \\
\cline{1-7}

IP
& T0$\rightarrow$T1
& +0.11
& -2.17
& +0.69
& +1.14
& +0.23 \\

&
T1$\rightarrow$T2
& -0.13
& \textbf{+0.53}
& \textbf{-7.14}
& +1.87
& \textbf{+4.87} \\

&
T2$\rightarrow$T3
& -0.05
& \textbf{-0.60}
& 0.00
& \textbf{+4.42}
& \textbf{-3.77} \\
\cline{1-7}

RU
& T0$\rightarrow$T1
& -0.36
& -0.72
& +0.67
& +0.06
& +0.35 \\

&
T1$\rightarrow$T2
& -0.89
& \textbf{+0.77}
& \textbf{-7.95}
& +1.75
& \textbf{+6.32} \\

&
T2$\rightarrow$T3
& \textbf{+0.30}
& -0.31
& +0.29
& \textbf{+4.33}
& \textbf{-4.61} \\
\cline{1-7}

OP
& T0$\rightarrow$T1
& +0.01
& -0.77
& +0.18
& +0.40
& +0.18 \\

&
T1$\rightarrow$T2
& -0.08
& +0.51
& \textbf{-7.30}
& \textbf{+3.27}
& \textbf{+3.60} \\

&
T2$\rightarrow$T3
& -0.03
& -0.27
& -0.76
& \textbf{+3.89}
& \textbf{-2.83} \\
\cline{1-7}

NP
& T0$\rightarrow$T1
& +0.02
& -0.47
& +0.04
& +0.31
& +0.10 \\

&
T1$\rightarrow$T2
& -0.05
& +0.30
& \textbf{-3.22}
& +0.92
& \textbf{+2.05} \\

&
T2$\rightarrow$T3
& -0.02
& -0.17
& -0.51
& \textbf{+2.33}
& \textbf{-1.63} \\
\hline
\end{tabular}
}

\caption{Mean within-user changes in relative attention across adjacent
transitions, in percentage points (pp). Population labels are All (Overall),
IP (Israel--Palestine), RU (Russia--Ukraine), OP (Other Political), and
NP (Non-Political). Displayed values use
sum-preserving rounding so that each row sums to zero.}
\label{tab:attention-share-shifts}
\end{table}

Tables~\ref{tab:attention-pattern-disruption-return} and
\ref{tab:attention-share-shifts} show that the conflict produced selective
rather than uniform disruption. Users with established Russia--Ukraine and
Israel--Palestine attention were the most consistently disrupted, while
Other Political users also showed substantial reorganization, and the Overall
and Non-Political populations were less uniformly affected. The share shifts
clarify the form of this disruption: during T1--T2, increased IUI attention
was accompanied primarily by a loss of Other Political attention, while
Israel--Palestine attention increased rather than being replaced; among
Israel--Palestine users in particular, the disruption combined increased IUI
and Israel--Palestine attention with a pronounced reduction in Other
Political attention. By T3, IUI attention declined while Non-Political
attention increased across populations. Direct comparison with the T1
reference further shows the strongest return toward pre-conflict attention
patterns among Russia--Ukraine and Israel--Palestine users, whereas the
Overall and Non-Political populations showed greater persistence, and Other
Political users exhibited a more mixed pattern. Thus, the conflict most
strongly reorganized users already engaged with other geopolitical conflicts,
and these same users were also the most likely to move back toward their
pre-conflict attention configuration.

\section{Discussion}
\label{sec:discussion}

The emergence of the Iran--US--Israel conflict did not produce a uniform shift in online attention. Instead, attention was reorganized through changes in participation, selective redistribution across topics, and uneven disruption of users' existing attention patterns. Together, the results qualify a simple zero-sum account of attention.
% The emergence of the Iran--US--Israel conflict reorganized attention through
% multiple mechanisms rather than a uniform shift toward a new topic. 
% Increased conflict-related engagement reflected reallocation
% among conflict-attentive users,
% activation of previously low-conflict-active users, and contraction
% among previously active users. Redistribution was
% also selective: Israel--Palestine was largely co-attended with the emerging conflict, while broader political and non-political content showed
% greater relative displacement. Disruption of
% attention patterns was more prevalent among users already attentive to geopolitical conflicts and partly reversed in the subsequent period. Together, these findings show that event-driven attention change involves changes in who participates, where attention is allocated, and how extensively
% existing attention patterns are disrupted. 

\noindent\textbf{Aggregate attention can conceal changes in participation.}
Aggregate increases in issue attention do not necessarily imply that an established public has collectively shifted its attention. The sharp rise in
Iran--US--Israel attention during T2 reflected both behavioral change among continuing participants and changes in the composition of the participating population. This matters because similar aggregate changes in issue prevalence or discussion \cite{stewart2020collective,smirnov2022covid,repke2024global} can arise from different user-level processes. From an agenda-setting and networked-gatekeeping perspective, sudden issue visibility may result not only from established politically engaged users changing what they amplify, but also from previously less conflict-active users entering conflict-related discussion. %Major events may therefore reshape online attention partly by changing the population contributing to an issue's visibility, rather than simply by redirecting the attention of an existing audience.

% Looking only at
% aggregate volume (Fig.~\ref{fig:topic-distribution}), the sharp rise in
% Iran--US--Israel attention during T2 could appear to reflect a collective shift toward
% the emerging conflict. The longitudinal analysis instead reveals that this aggregate increase reflected 
% % substantial turnover in the composition and behavior of the participating population. 
% both behavioral change among continuing participants and a change in the composition of the participating population. 
% The
% same aggregate increase in topic volume can therefore arise without an established audience
% collectively switching its attention.
% This distinction matters for
% event-attention research based primarily on issue prevalence or discussion volume over
% time \cite{stewart2020collective,smirnov2022covid,repke2024global}. From an
% agenda-setting and networked-gatekeeping perspective, sudden issue visibility
% may arise not only because established politically engaged users change what
% they amplify, but also because an event draws previously less conflict-active users into conflict-related amplification.
% mobilizes new amplifiers. Thus, the
% same aggregate increase in topic volume can arise from changes in the
% composition of participating users rather than an established audience
% collectively switching attention.

% The Finite Pool of Attention hypothesis proposes that attending more to one
% threat reduces the attention available for others 

\noindent\textbf{Attention expansion and competition can coexist.}
The Finite Pool of Attention perspective suggests that increased attention to one issue reduces the attention available for others \cite{sisco2023finite}. Our results qualify a simple zero-sum interpretation. In absolute terms, increased Iran--US--Israel attention often coincided with increased activity toward existing topics, indicating that users could temporarily expand their overall repost activity even while relative attention shifted substantially across topics. Since relative shares are compositionally constrained, the key question is not whether some share declines as Iran--US-Israel attention increases, but rather where the displacement occurs. Our results show that displacement was selective. Israel--Palestine largely maintained its relative prominence alongside the emerging conflict, while broader political and non-political content experienced greater displacement, with Russia--Ukraine showing weaker, more heterogeneous displacement. Rather than contradicting the Finite Pool of Attention perspective, our findings suggest that observable engagement can expand even as competition for relative attention remains selective across topics.

\noindent\textbf{Related issues may be co-attended rather than displaced.}
The contrast between Israel--Palestine and Russia--Ukraine suggests that relationships among issues may shape how attention is reorganized. Israel--Palestine, which shares a central actor and regional context with Iran--US--Israel, was strongly co-attended with the emerging conflict. The Russia–Ukraine conflict, by contrast, showed greater relative displacement, and users focused on it were also more likely to shift directly toward an Iran–US–Israel-dominant profile. Israel--Palestine-focused users more often retained their existing focus or combined it with attention to the new conflict. These patterns are consistent with closely connected issues being bundled within the same attention repertoire rather than competing directly for prominence \cite{sisco2023finite,rauchfleisch2023covid}. The audience networks point in the same direction. Previously differentiated Russia--Ukraine- and Israel--Palestine-focused audiences became more interconnected around Iran--US--Israel-focused profiles during T2, with reduced within-profile linking and greater linking to Iran--US--Israel-related profiles. Although these patterns partially reversed in T3, they did not fully return to their earlier configuration, suggesting that major cross-cutting events may temporarily bridge previously differentiated conflict-attention audiences. With only one closely related and one more distant comparison, however, these results suggest rather than establish that topic relatedness fosters co-attention. 
% Broader comparisons across simultaneous events are needed to test this directly in future work.

\noindent\textbf{Attention shocks are uneven and partly reversible.}
The emerging conflict did not constitute a uniform platform-wide attention shock. Disruption was substantially more prevalent among users already attentive to geopolitical conflicts than among the broader or primarily Non-Political populations, indicating that the event most strongly reorganized users for whom geopolitical issues were already part of their attention repertoire. Yet stronger disruption did not necessarily imply more persistent change. Russia--Ukraine and Israel--Palestine users, who showed the clearest conflict period reorganization, were also the groups most likely to move back to their pre-conflict attention pattern by T3. This distinction between disruption and persistence suggests that the users most affected when a crisis emerges are not necessarily those whose attention patterns remain most durably altered.

% Assessing disruption therefore requires considering not only its average
% magnitude but also its prevalence across users. Russia--Ukraine and Other
% Political users, for example, showed relatively similar average disruption,
% yet disruption relative to the pre-conflict transition occurred among a
% substantially larger share of Russia--Ukraine users. Average disruption alone
% can therefore obscure how consistently an event affects members of a
% population. The conflict thus most strongly and consistently reorganized
% users for whom geopolitical conflicts were already part of their attention
% repertoire, rather than producing uniform disruption across the platform.

\subsubsection*{Limitations.} %and Future Work}
Our findings should be interpreted in light of several limitations. Reposting captures active amplification rather than passive exposure or attention outside Bluesky. Our event-centered design identifies changes around conflict onset rather than a causal effect of the conflict itself. Our analysis is also limited to English-language Bluesky activity, which may limit generalizability to other platforms and languages. For the attention patterns analysis, we focus on users who are active in all four windows, which improves within-user comparability but excludes more transient participants. Our five content categories and three focal conflicts simplify the broader information environment. Finally, our approximately three-month windows capture sustained changes but may conceal short-lived responses around conflict onset. Future work could address these limitations through finer temporal resolution, cross-platform and multilingual comparisons, and analyses of a broader set of overlapping events.

%\subsubsection*{Ethical statement.}
%We use public and published data from Bluesky following the terms of service and privacy regulations. While the raw data contains personally identifiable information, all our analysis is based on aggregated results where individuals cannot be identified.
%Our findings could in principle inform efforts to divert attention from a conflict, but they characterize population-level phenomena rather than interventions, and their main value lies in helping to understand how attention to ongoing crises is sustained or lost.

% Bluesky's population and platform norms may limit generalizability to other
% platforms, while our focus on English-language content and active reposters
% captures only part of users' attention. Reposting measures active amplification
% rather than passive exposure, liking, or off-platform attention. Our five
% content categories and three focal conflicts simplify the broader information
% environment; future work could examine more events to test systematically how
% topic relatedness shapes competition and co-attention. Finally, our
% approximately three-month windows capture sustained changes but may conceal
% short-lived responses around conflict onset.

\section{Conclusion}
\label{sec:conclusion}

Major attention shocks cannot be understood from aggregate volume alone. By following the same users across an emerging crisis, we show that apparent shifts in collective attention can reflect changes in who participates, selective competition and co-attention across topics, and disruptions that differ both in magnitude and persistence across users. Understanding event-driven attention therefore requires examining not only how much attention an event receives, but who supplies it, where it comes from, and how it reorganizes existing patterns of engagement. 

\section{Acknowledgments}

This work was supported by the Research Council of Finland project
\textit{POLEMIC: POLarization Entangled with Mental Well-being: Integrated and
Computational Approach to Quantifying the Underlying Feedback Loop}
(Decision No.~371535, 371536), at Aalto University.
We wish to acknowledge the Aalto University Science-IT project for generous
computational resources.

% Major events reorganize online attention through changes in who participates,
% where attention moves, and how existing attention patterns are disrupted.
% Aggregate volume can conceal population turnover, while absolute expansion can
% coexist with relative competition and that competition can vary across topics.
% Disruption likewise depends on users' prior attention repertoires rather than
% affecting the platform uniformly. Understanding event-driven attention
% therefore requires examining not only how much attention an event receives,
% but who supplies it, where it comes from, and how it reorganizes existing
% patterns of engagement.

\bibliographystyle{unsrtnat}
\bibliography{aaai2026}

\appendix

\section{Collection Pipeline and Coverage}
\label{app:collection}

\subsection{Collection Pipeline}

Data were collected via a live stream from the Bluesky Firehose, an authenticated real-time stream of events encoded according to the AT Protocol. The collection pipeline decoded incoming CBOR events and stored them as raw Newline Delimited JSON (NDJSON) files. While the pipeline captures all available event types, including posts, likes, and follows, this study exclusively filtered and utilized repost events involving English-language posts. English-language content was identified using the language metadata associated with the original post. The collection architecture ensured real-time data capture with automatic partition management for efficient querying.

\subsection{Reliability and Coverage}

The broader collection period contained several interruptions or incomplete
days. No dates were excluded from T0 or T3, one date (21 February 2026) was
excluded from T1, and four dates (7 March, 14 March, 25 April, and 2 May 2026)
were excluded from T2. Consequently, the analyses cover 90 days in T0, 89 days
in T1, 86 days in T2, and 90 days in T3.

\section{Topic Classification Prompt}
\label{app:classification-prompt}

Posts were assigned to one of five mutually exclusive content categories using the classification prompt shown in Listing~\ref{lst:classification-prompt}.

\begin{lstlisting}[
caption={Prompt for Conflict Topic Classification},
label={lst:classification-prompt},
basicstyle=\ttfamily\footnotesize,
breaklines=true,
breakatwhitespace=false,
columns=fullflexible
]
You are a social media post classifier.
Your task is to read a post and assign it to exactly one category from the list below.

# Categories

### Russia-Ukraine Conflict
Posts about conflict, war, escalation, diplomacy, or policy directly related to Russia and Ukraine.
Covers references to Russia, Ukraine, Kyiv, Kremlin, Zelensky, Putin, Donbas, Crimea, NATO, invasion, territorial disputes, military aid, sanctions, ceasefires, casualties, humanitarian impacts, peace talks, or official government/organization statements related to the Russia-Ukraine conflict.
Use this category only when the post clearly connects Russia, Ukraine, or related actors to the war, military action, territorial conflict, sanctions, aid, diplomacy, or escalation.
Do not use this category for general mentions of Russia, Ukraine, Putin, or Zelensky that are not clearly about the conflict.

### Israel-Palestine Conflict
Posts about conflict, occupation, escalation, diplomacy, or policy directly related to Israel and Palestine.
Covers references to Israel, Palestine, Gaza, West Bank, Hamas, Netanyahu, IDF, settlers, occupation, UNRWA, airstrikes, hostages, ceasefires, casualties, humanitarian impacts, sanctions, aid, peace talks, or official government/organization statements related to the Israel-Palestine conflict.
Use this category only when the post clearly connects Israel, Palestine, Gaza, the West Bank, or related actors to conflict, occupation, military action, sanctions, aid, diplomacy, or escalation.
Do not use this category for general mentions of Israel, Palestine, Netanyahu, or Hamas that are not clearly about the conflict.

### Iran-US-Israel Conflict
Posts about conflict, escalation, diplomacy, or high-stakes geopolitical confrontation involving Iran, the United States, and/or Israel.
Covers references to Iran, Tehran, IRGC, Israel, Jerusalem, the United States, Washington, U.S. forces, proxy groups, nuclear tensions, missile or drone attacks, airstrikes, retaliation, sanctions, military deployments, ceasefires, casualties, humanitarian impacts, diplomacy, or official government/organization statements related to Iran-US-Israel tensions.
Use this category only when the post clearly connects Iran, the United States, Israel, or related actors to conflict, military action, escalation, sanctions, nuclear tensions, proxy warfare, diplomacy, or regional security tensions.
Do not use this category for general mentions of Iran, the United States, Israel, Trump, Biden, Khamenei, or Netanyahu that are not clearly about conflict or geopolitical escalation.

### Other Political
Posts that are political, geopolitical, governmental, legal, electoral, activist, or public-policy related but do not fit any specific conflict category above.
Use this category for general country/politician mentions when the post is political but not clearly about one of the three conflicts.

### Non-Political
Posts not related to politics, public policy, government, war, diplomacy, or activism.
Includes entertainment, sports, personal life, culture, memes, gaming, marketing, lifestyle, and other non-political content.

# Classification rules

- Assign the post to exactly one category.
- Prefer the most specific applicable category, but only when the post clearly matches that category.
- Use only the content of the post itself. Hashtags, mentions, linked text, and emojis count as content.
- Do not infer unstated facts or background context.
- If a post clearly relates to multiple conflict categories, choose the category that is the primary focus of the text.
- If a post is political but does not clearly fit one of the three conflict categories, use Other Political.
- If a post is not political, use Non-Political.
\end{lstlisting}

\section{Conflict-Attention Profile Mixing Matrices}
\label{app:profile-mixing}

To quantify the organization of conflict-focused repost audiences, we
calculated the distribution of outgoing repost links from each source
conflict-attention profile across target profiles. For source profile $i$ and
target profile $j$, the mixing probability is

\[
P(j \mid i)
=
\frac{E_{i\rightarrow j}}
{\sum_k E_{i\rightarrow k}},
\]

where $E_{i\rightarrow j}$ is the number of retained directed repost edges
from users in source profile $i$ to users in target profile $j$. Each row is
therefore normalized separately and sums to 100\%. Table~\ref{tab:profile-mixing-matrices}
reports these distributions for T1--T3.

\begin{table*}[!t]
\centering
\small
\setlength{\tabcolsep}{2.5mm}

\begin{tabular}{@{}lrrrrrrr@{}}
\hline
\multicolumn{8}{c}{\textbf{T1}} \\
\hline
\textbf{Source $\backslash$ Target}
& \textbf{RU}
& \textbf{IP}
& \textbf{IUI}
& \textbf{IUI+IP}
& \textbf{IUI+RU}
& \textbf{IP+RU}
& \textbf{All Three} \\
\hline
RU        & \textbf{87.02} & 1.77 & 4.07 & 0.07 & 5.53 & 0.47 & 1.08 \\
IP        & 1.17 & \textbf{93.71} & 0.74 & 1.49 & 0.55 & 0.81 & 1.54 \\
IUI       & 12.64 & 4.71 & \textbf{69.64} & 0.87 & 9.67 & 0.62 & 1.86 \\
IUI+IP    & 4.35 & \textbf{65.94} & 6.84 & 5.75 & 5.91 & 2.02 & 9.18 \\
IUI+RU    & \textbf{43.67} & 4.14 & 19.29 & 0.48 & 25.99 & 1.01 & 5.42 \\
IP+RU     & 39.75 & \textbf{43.58} & 3.57 & 0.94 & 5.48 & 2.54 & 4.14 \\
All Three & \textbf{32.62} & 30.25 & 9.31 & 1.59 & 15.91 & 2.23 & 8.09 \\
\hline
\end{tabular}

\medskip

\begin{tabular}{@{}lrrrrrrr@{}}
\hline
\multicolumn{8}{c}{\textbf{T2}} \\
\hline
\textbf{Source $\backslash$ Target}
& \textbf{RU}
& \textbf{IP}
& \textbf{IUI}
& \textbf{IUI+IP}
& \textbf{IUI+RU}
& \textbf{IP+RU}
& \textbf{All Three} \\
\hline
RU        & \textbf{76.80} & 0.22 & 13.55 & 0.63 & 7.90 & 0.00 & 0.89 \\
IP        & 0.16 & \textbf{63.30} & 9.37 & 26.09 & 0.22 & 0.01 & 0.85 \\
IUI       & 5.65 & 2.92 & \textbf{81.29} & 5.13 & 2.75 & 0.01 & 2.25 \\
IUI+IP    & 1.67 & 32.63 & \textbf{32.71} & 29.88 & 1.01 & 0.05 & 2.07 \\
IUI+RU    & 35.21 & 2.38 & \textbf{48.94} & 3.98 & 7.17 & 0.01 & 2.30 \\
IP+RU     & \textbf{43.85} & 25.38 & 6.92 & 15.38 & 5.38 & 0.00 & 3.08 \\
All Three & 22.37 & 13.30 & \textbf{41.75} & 15.03 & 4.22 & 0.02 & 3.32 \\
\hline
\end{tabular}

\medskip

\begin{tabular}{@{}lrrrrrrr@{}}
\hline
\multicolumn{8}{c}{\textbf{T3}} \\
\hline
\textbf{Source $\backslash$ Target}
& \textbf{RU}
& \textbf{IP}
& \textbf{IUI}
& \textbf{IUI+IP}
& \textbf{IUI+RU}
& \textbf{IP+RU}
& \textbf{All Three} \\
\hline
RU        & \textbf{85.01} & 2.04 & 7.85 & 0.33 & 4.43 & 0.12 & 0.22 \\
IP        & 1.45 & \textbf{88.41} & 3.52 & 4.34 & 1.09 & 0.51 & 0.68 \\
IUI       & 9.32 & 8.62 & \textbf{72.30} & 2.14 & 6.66 & 0.09 & 0.87 \\
IUI+IP    & 6.72 & \textbf{45.74} & 33.29 & 7.56 & 4.84 & 0.35 & 1.50 \\
IUI+RU    & 36.34 & 8.33 & \textbf{43.24} & 1.88 & 9.17 & 0.16 & 0.88 \\
IP+RU     & \textbf{45.63} & 40.24 & 5.89 & 2.54 & 3.35 & 1.73 & 0.61 \\
All Three & 26.48 & 27.34 & \textbf{32.34} & 3.92 & 7.96 & 0.47 & 1.48 \\
\hline
\end{tabular}

\caption{Row-normalized mixing of directed conflict-related repost links
between conflict-attention profiles across T1--T3. Each cell reports the
percentage of retained outgoing repost links from users in the source profile
(row) that are directed toward users in the target profile (column); each row
therefore sums to 100\% up to rounding. Bold values indicate the largest
target share within each source-profile row. RU denotes Russia--Ukraine, IP
denotes Israel--Palestine, and IUI denotes Iran--US--Israel.}
\label{tab:profile-mixing-matrices}
\end{table*}

\section{Non-IUI Baseline Analysis}
\label{app:non_iui_baseline}

No additional category-specific restriction was imposed for the aggregated
non-IUI comparison. For the aggregated non-IUI comparison, absolute non-IUI attention was defined as
\[
n_{u,t}^{\mathrm{non\text{-}IUI}}
=
N_{u,t} - n_{u,t}^{\mathrm{IUI}},
\]
such that
\[
\Delta_{u,\mathrm{non\text{-}IUI}}^{\mathrm{abs}}
=
\left(
N_{u,T2} - n_{u,T2}^{\mathrm{IUI}}
\right)
-
\left(
N_{u,T1} - n_{u,T1}^{\mathrm{IUI}}
\right).
\]

This absolute relationship is not mechanically constrained because users'
total repost activity may change between T1 and T2. By contrast, the relative
non-IUI measure is the complement of the Iran--US--Israel attention share in
each window,
\[
a_{u,t}^{\mathrm{non\text{-}IUI}}
=
1 - a_{u,t}^{\mathrm{IUI}},
\]
and therefore
\[
\Delta_{u,\mathrm{non\text{-}IUI}}^{\mathrm{rel}}
=
-\Delta_{u,\mathrm{IUI}}^{\mathrm{rel}}.
\]

The relative Iran--US--Israel versus non-IUI relationship is consequently
deterministic rather than an empirical test of selective competition, and we
treat it as a compositional baseline. This pattern is already visible in the aggregated non-IUI baseline: absolute
IUI and non-IUI changes were positively correlated in every T1 behavioral
group, with Pearson correlations ranging from approximately $r=.10$ to
$r=.87$. Thus, at the level of absolute repost counts, the emergence of the
Iran--US--Israel conflict was accompanied predominantly by positive
co-movement with other activity rather than one-for-one replacement.

\begin{table*}[t]
\centering
\small
\setlength{\tabcolsep}{1mm}

\begin{tabular}{ccccc}
\hline
\multicolumn{5}{c}{\textbf{Panel A: Conflict and non-IUI comparisons}} \\
\hline
\textbf{Political attention} &
\textbf{Activity} &
\textbf{Non-IUI} &
\textbf{Israel--Palestine} &
\textbf{Russia--Ukraine} \\
\hline

$[0,.3)$ & 1--4
& 257,828 / .371 / -1.000
& 270 / .958 / -.066
& 82 / .756 / -.177 \\

$[0,.3)$ & 5--14
& 171,855 / .218 / -1.000
& 474 / .809 / -.001
& 110 / .586 / -.137 \\

$[0,.3)$ & 15--34
& 104,652 / .095 / -1.000
& 1,006 / .891 / .056
& 141 / .145 / -.029 \\

$[0,.3)$ & 35--69
& 65,501 / .157 / -1.000
& 1,544 / .499 / .131
& 216 / .199 / .155 \\

$[0,.3)$ & 70+
& 122,559 / .149 / -1.000
& 17,580 / .276 / .221
& 3,371 / .337 / .048 \\

\hline

$[.3,.6)$ & 1--4
& 36,775 / .593 / -1.000
& 280 / .657 / -.027
& 143 / .944 / -.242 \\

$[.3,.6)$ & 5--14
& 38,705 / .875 / -1.000
& 915 / .339 / -.025
& 458 / .300 / -.055 \\

$[.3,.6)$ & 15--34
& 26,085 / .547 / -1.000
& 1,801 / .607 / .047
& 854 / .336 / -.027 \\

$[.3,.6)$ & 35--69
& 16,787 / .525 / -1.000
& 2,468 / .887 / .099
& 1,072 / .472 / -.065 \\

$[.3,.6)$ & 70+
& 35,468 / .168 / -1.000
& 18,147 / .539 / .142
& 11,138 / -.016 / .045 \\

\hline

$[.6,1]$ & 1--4
& 76,356 / .734 / -1.000
& 943 / .442 / -.133
& 964 / .875 / -.163 \\

$[.6,1]$ & 5--14
& 62,419 / .525 / -1.000
& 2,532 / .918 / -.124
& 2,674 / .410 / -.138 \\

$[.6,1]$ & 15--34
& 43,846 / .849 / -1.000
& 3,751 / .887 / -.095
& 4,112 / .753 / -.124 \\

$[.6,1]$ & 35--69
& 31,889 / .739 / -1.000
& 4,645 / .690 / -.032
& 5,574 / .378 / -.082 \\

$[.6,1]$ & 70+
& 92,686 / .102 / -1.000
& 48,695 / .515 / .015
& 58,923 / .100 / -.066 \\

\hline
\end{tabular}

\medskip

\begin{tabular}{cccc}
\hline
\multicolumn{4}{c}{\textbf{Panel B: Other Political and Non-Political comparisons}} \\
\hline
\textbf{Political attention} &
\textbf{Activity} &
\textbf{Other Political} &
\textbf{Non-Political} \\
\hline

$[0,.3)$ & 1--4
& 9,324 / .881 / -.075
& 157,389 / .323 / -.292 \\

$[0,.3)$ & 5--14
& 21,693 / .488 / .002
& 171,855 / .152 / -.280 \\

$[0,.3)$ & 15--34
& 29,914 / .692 / .033
& 104,652 / .051 / -.296 \\

$[0,.3)$ & 35--69
& 26,222 / .417 / .029
& 65,501 / .121 / -.271 \\

$[0,.3)$ & 70+
& 76,362 / .053 / .013
& 122,559 / .144 / -.251 \\

\hline

$[.3,.6)$ & 1--4
& 12,543 / .902 / -.171
& 18,237 / .637 / -.246 \\

$[.3,.6)$ & 5--14
& 35,359 / .919 / -.105
& 38,705 / .688 / -.259 \\

$[.3,.6)$ & 15--34
& 26,026 / .622 / -.064
& 26,085 / .395 / -.286 \\

$[.3,.6)$ & 35--69
& 16,767 / .536 / -.034
& 16,787 / .344 / -.322 \\

$[.3,.6)$ & 70+
& 35,460 / -.021 / -.018
& 35,468 / .270 / -.330 \\

\hline

$[.6,1]$ & 1--4
& 39,980 / .723 / -.275
& 13,496 / .619 / -.211 \\

$[.6,1]$ & 5--14
& 61,660 / .857 / -.286
& 40,174 / .726 / -.200 \\

$[.6,1]$ & 15--34
& 43,766 / .828 / -.320
& 41,613 / .820 / -.197 \\

$[.6,1]$ & 35--69
& 31,863 / .742 / -.346
& 31,709 / .549 / -.188 \\

$[.6,1]$ & 70+
& 92,666 / -.004 / -.388
& 92,657 / .238 / -.224 \\

\hline
\end{tabular}

\caption{Population sizes and Pearson correlations between
Iran--US--Israel attention changes and comparison-category attention changes
across T1 behavioral groups. Each comparison cell reports
$n / r_{\mathrm{abs}} / r_{\mathrm{rel}}$, where $n$ is the number of
users, $r_{\mathrm{abs}}$ is the Pearson correlation between absolute
attention changes, and $r_{\mathrm{rel}}$ is the Pearson correlation between
relative attention changes. Activity denotes total annotated repost activity
in T1, and political attention denotes the share of T1 attention devoted to
political content. For the non-IUI baseline, $r_{\mathrm{rel}}=-1$ by
construction because the non-IUI share is the complement of the
Iran--US--Israel share.}
\label{tab:attention-correlations-by-group}
\end{table*}

\section{Validation of LLM-Generated Topic Labels}
\label{app:llm-validation}

To assess the reliability of the LLM-generated topic classifications, we
constructed a stratified random validation sample based on the label assigned
by the LLM. The classification scheme consisted of five mutually exclusive
categories: \textit{Iran--US--Israel Conflict},
\textit{Israel--Palestine Conflict}, \textit{Russia--Ukraine Conflict},
\textit{Other Political}, and \textit{Non-Political}. Sampling was stratified
by the LLM-assigned label so that relatively rare conflict-related categories
were sufficiently represented in the validation sample. Each sampled post was
then manually assigned a topic label, with the human classification treated as
the reference label. The 239 sampled posts were each manually coded by a single annotator with substantial familiarity with the relevant geopolitical contexts; each post received a single annotation. The annotator was
blinded to the LLM-assigned labels.

Because sampling was conditional on the LLM-assigned category, our primary
category-level validation measure is class-specific precision, defined as
\[
\Pr(\text{Human label}=k \mid \text{LLM label}=k).
\]
This quantity measures the proportion of posts assigned to category $k$ by the
LLM that received the same category under human coding. We report 95\%
Wilson confidence intervals for these class-specific proportions. The
macro-average precision, which assigns equal weight to each of the five
categories, was 95.4\% (95\% bootstrap CI: 92.4--97.9\%).

Since the validation sample deliberately contained a more balanced
representation of LLM labels than the full corpus, the unweighted validation
sample does not directly estimate overall corpus-level accuracy. We therefore
reweighted observations according to the frequency of each LLM-assigned label
in the full corpus. This procedure yielded an estimated population-wide
classification accuracy of 97.1\% (95\% stratified-bootstrap CI:
92.7--99.8\%). Bootstrap confidence intervals were obtained by resampling
observations within each LLM-label stratum while holding the full-corpus
stratum sizes fixed. Overall, the validation results indicate a high degree of
correspondence between the automated and human topic classifications.

Table~\ref{tab:llm_validation} summarizes the validation results by category.

\begin{table}[!htbp]
\centering
\small
\setlength{\tabcolsep}{2.5pt}
\renewcommand{\arraystretch}{1.0}

\begin{tabular}{@{}lrrrr@{}}
\hline
\textbf{LLM-assigned label}
& \textbf{$N$}
& \textbf{Agree}
& \textbf{Prec.}
& \textbf{95\% CI} \\
\hline
Iran--US--Israel
& 38 & 33 & 86.8\% & [72.7, 94.2] \\
Israel--Palestine
& 52 & 52 & 100.0\% & [93.1, 100.0] \\
Russia--Ukraine
& 49 & 47 & 95.9\% & [86.3, 98.9] \\
Other Political
& 61 & 59 & 96.7\% & [88.8, 99.1] \\
Non-Political
& 39 & 38 & 97.4\% & [86.8, 99.5] \\
\hline
\textbf{Macro average}
& --- & --- & \textbf{95.4\%} & \textbf{[92.4, 97.9]} \\
\textbf{Pop.-weighted acc.}
& --- & --- & \textbf{97.1\%} & \textbf{[92.7, 99.8]} \\
\hline
\end{tabular}

\caption{Agreement between LLM-generated and human topic labels.
Category-level confidence intervals are 95\% Wilson intervals; macro-average
and population-weighted intervals are based on stratified bootstrap
resampling.}
\label{tab:llm_validation}
\end{table}

Class-specific precision was high across all five labels, ranging from 86.8\%
for the Iran--US--Israel category to 100.0\% for the Israel--Palestine
category. Category-specific confidence intervals are 95\% Wilson intervals,
while confidence intervals for the macro average and population-weighted
accuracy are based on 50000 stratified bootstrap resampling.

\end{document}